\documentclass{article}

\usepackage{blindtext}
\usepackage{lipsum}

\let\OLDthebibliography\thebibliography
\renewcommand\thebibliography[1]{
  \OLDthebibliography{#1}
  \setlength{\parskip}{1pt}
  \setlength{\itemsep}{3pt plus 0.3ex}
}
\usepackage{hyperref}
\usepackage{amssymb}
\usepackage{amsmath}
\usepackage{amsthm}
\usepackage{mathrsfs}
\usepackage[arrow,matrix,curve]{xy}
\usepackage{graphicx}
\usepackage{hyperref}
\usepackage{amscd,verbatim}
\usepackage{mathtools}
\usepackage{tikz-cd}
\usepackage{comment}
\usepackage[affil-it]{authblk}

\usepackage{fullpage,enumerate}

\usepackage{fullpage,enumerate}
\newcommand{\bea}{\begin{eqnarray}}
\newcommand{\eea}{\end{eqnarray}}
\newcommand{\beq}{\begin{equation}}
\newcommand{\eeq}{\end{equation}}

\newcommand\wt{\widetilde}

\theoremstyle{remark}

\numberwithin{theorem}{section}

\newcommand{\TT}{{\mathbb T}}
\newcommand{\ZZ}{{\mathbb Z}}

\begin{document}

\title{\bf Global topology of Weyl semimetals and Fermi arcs}

\author{Varghese Mathai%
  \thanks{Electronic address: \texttt{mathai.varghese@adelaide.edu.au}}\,\,}
\author{Guo Chuan Thiang%
  \thanks{Electronic address: \texttt{guochuan.thiang@adelaide.edu.au}}}
\affil{School of Mathematical Sciences, University of Adelaide, Adelaide, SA 5005, Australia
 }
  
  \date{}





\maketitle
\vspace{-.3in}
\begin{abstract}
We provide a manifestly topological classification scheme for generalised Weyl semimetals, in any spatial dimension and with arbitrary Weyl surfaces which may be non-trivially linked. The classification naturally incorporates that of Chern insulators. Our analysis refines, in a mathematically precise sense, some well-known 3D constructions to account for subtle but important global aspects of the topology of semimetals. Using a fundamental locality principle, we derive a generalized charge cancellation condition for the Weyl surface components. We analyse the bulk-boundary correspondence under a duality transformation, which reveals explicitly the topological nature of the resulting surface Fermi arcs. We also analyse the effect of moving Weyl points on the bulk and boundary topological semimetal invariants.\\
\end{abstract}


\section{Introduction}
3D Weyl semimetals have attracted much attention recently, due to the possibility of realising elusive ``Weyl fermions'' in condensed matter systems \cite{Wan, Turner, Weng}. The experimental signature of a Weyl semimetal was predicted to be \emph{Fermi arcs} of gapless surface modes. Fermi arcs have recently been experimentally found \cite{Xu1, Xu2, Lv, Zhang}, as have the related Fermi nodal lines \cite{Bian} and Weyl nodes \cite{Lv2}, prompting intense research into the general theory of semimetals \cite{Turner, Zhao1, Zhao2} including those in five spatial dimensions \cite{ZhangLian}.
A Weyl semimetal can be thought of as an insulator with band crossings at isolated ``Weyl points'' in momentum space, whose behaviour parallels that of magnetic monopoles. Its stability has a topological origin --- surround a Weyl point by a sphere on which the Fermi projection is well-defined, then calculate the integrated Berry curvature of the valence bands over this sphere to obtain a quantised Chern number (the monopole charge/chirality of the Weyl point). The stability of the corresponding surface Fermi arcs may thereby be inferred from a bulk-boundary correspondence, but it is desirable to understand their topological nature directly and in a rigorous mathematical framework. \\

{\bf Overview of new results.} One of our contributions in this Letter is the identification of Fermi arc invariants with elements of a \emph{relative homology group}. We argue that the local charges of the Weyl points do not unambiguously characterise the topological class of Fermi arcs or semimetal band structures --- there may be globally inequivalent band structures/Fermi arcs which nevertheless have the same charges at the Weyl points. Indeed, the spheres around the Weyl points only see a local portion of the full bulk band topology, and some complementary data is needed to encode how the insulating band structure is extended globally from these spheres to the rest of the Brillouin zone. The assembly of these data into our more refined classification group follows from a well-known locality principle (the \emph{Mayer--Vietoris principle} \cite{BottTu}) in topology. Our topological invariants \emph{simultaneously} incorporate the usual classifications of topological insulators (via Chern numbers etc.) and semimetals (via the Weyl point charges). In fact, we show that the insulator invariants and Weyl charge invariants interact inside a larger group of semimetal invariants which is able to capture, for instance, the standard two-level Weyl semimetal augmented with additional (possibly non-trivial) valence bands. Indeed, the local Weyl point charges exactly capture the topological obstruction to gapping out band crossings in a semimetal to form a (possibly nontrivial topological) insulator. Our general analysis is important because additional non-trivial valence bands can modify the Fermi arc topology (Fig.\ \ref{fig:ambiguity}).

Usually, one allows perturbations of the band structure which may move the Weyl points around slightly inside the Brillouin torus without changing their local charges. A careful analysis involves studying the \emph{global implementation} of such perturbations, and we provide examples where semimetal band structures/Fermi arcs can look the same locally around the Weyl points, but cannot be transformed into one another by continuously moving the (projected) Weyl points along with the Fermi arcs inside the (projected) Brillouin torus (Fig.\ \ref{fig:ambiguity2}-\ref{fig:ambiguity3}). The required global perturbation in such cases is ``large'', and motivates us to study general perturbations up to \emph{ambient isotopy}, and their effect on the topological classification of semimetals.

\begin{figure}[ht]
\begin{center}
\includegraphics[scale=1]{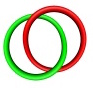}\qquad\includegraphics[scale=1]{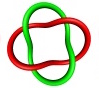}
\end{center}
\caption{\small\it Disjoint tubular neighbourhoods for some well known links.}
\label{fig:toruslink}
\end{figure}

In higher dimensions, one generally has Weyl surfaces instead of points, with potentially interesting physical consequences \cite{ZhangLian}. For instance, the ``enveloping sphere'' construction does not work for \emph{linked} loops, and the appropriate way to surround each loop disjointly (to establish their individual topological charges) is to use \emph{tubular neighbourhoods} (Fig.\ \ref{fig:toruslink}). A sphere surrounding the entire link sees only the total enclosed charge and loses significant topological information about the Weyl surfaces. Interestingly, a Stokes' theorem argument constrains the total charge of the Weyl points of a semimetal to be zero \cite{Witten}, and we show how this follows more generally from the Mayer--Vietoris principle: this is a global condition inherited from the \emph{global existence} of a semimetal band structure.\\


\section{Topological semimetal phases and cohomology}
{\bf Topological band insulators and cohomology groups.}
The 2D Chern insulator is classified by a (first) Chern number which is equal to the integrated Berry curvature 2-form of the valence bands over the Brillouin zone $T$. The Chern number captures the topological non-triviality of the insulator in the sense of a topological obstruction to globally defining a Berry phase for the valence band. Formally, this classification problem is equivalent to that of U(1) line bundles over the Brillouin zone, which are classified by the \emph{integral cohomology group} $H^2(\TT^2,\ZZ)\cong\ZZ$. In terms of differential forms, $H^2(\TT^2,\ZZ)\neq 0$ says that there are closed integral 2-forms $\mathcal{F}$ on $\TT^2$ (Berry curvatures) which are not globally exact $(\mathcal{F}\neq d\mathcal{A})$; generally, a degree-$n$ cohomology group $H^n(X,\ZZ)$ classifies closed integral $n$-forms on $X$ modulo the exact forms.

More complicated invariants classify topological insulators in higher dimensions which may also have additional symmetries (e.g.\ time-reversal or point group symmetries). For example, there are higher degree Chern numbers in 4D due to $H^4(\TT^4,\ZZ)\cong\ZZ$, and these are responsible for the generalised quantum Hall effect in 4D \cite{Zhang4D} (c.f.\ the TKNN treatment of the 2D QHE \cite{TKNN}). Topologically non-trivial insulating phases therefore have a \emph{cohomological} origin, owing to the fact that the Brillouin torus hosts non-trivial cohomology groups in which Chern numbers and their generalisations (e.g.\ spin-Chern numbers, Kane--Mele invariants) live. There are also $K$-theory classification schemes for topological insulators \cite{Kitaev, Thiang, MT2} and Fermi surfaces \cite{Horava, Zhao1, Zhao2} which are similar in spirit insofar as $K$-theory is a \emph{generalized cohomology} theory.\\

{\bf Weyl semimetals.}
In general, the Fermi level may pass through band crossings at some Weyl surfaces in momentum space, rather than lie completely inside a band gap. Such crossings may arise as ``accidental degeneracies'' \cite{vN, Herring} and, in 3D, occur generically at points. A quick way to see this is to consider two-level Bloch Hamiltonians $H(k)=\vec{h}(k)\cdot\vec{\sigma}$ parametrised by $k\in T$; here $\vec{h}(k)$ is a 3-vector, $\vec{\sigma}$ is the vector of Pauli matrices, and we have left out a constant energy term (setting the Fermi energy to 0). The Bloch eigenvalues are $\pm |\vec{h}(k)|$, so a band crossing at $k$ requires three conditions $\vec{h}(k)=\vec{0}$, and we see that the locus $W$ of band crossings generically has codimension 3. For $T=\TT^3$, we get a collection of Weyl points, and the low-energy excitations are essentially described by the Weyl equation. Furthermore, each Weyl point serves as a ``monopole'' for the Berry curvature, and may be assigned a topologically protected ``chirality/charge'' by integrating the Berry curvature over a small enclosing sphere (over which the valence band is well-defined).\\


{\bf Intuition: removing points affects topology significantly.} It is well known that removing points from a manifold can drastically modify its topological properties. A simple example is the 3-sphere $S^3$, which has trivial second homotopy group, $\pi_2(S^3)=0$. By removing even a single point $\{w\}$, the second homotopy group becomes non-zero --- the map from $S^2$ to a small sphere in $S^3$ surrounding $\{w\}$ is no longer null-homotopic. It is this principle which accounts for the possibility of new topological semimetal phases, for which the Fermi projection need only be well-defined on the Brillouin torus with some Weyl points removed. The precise change in topology when points are removed also depends strongly on the ambient manifold, which needs to be taken into account in the analysis. As we explain later, the second cohomology group of the 3-torus $\TT^3$ with \emph{two} points removed has an extra copy of $\ZZ$ compared with that of $\TT^3$, and this extra subgroup accounts for the possibility of the basic Weyl semimetal phase in 3D. \\

{\bf The Weyl charge group}
A general mathematical framework for classifying Weyl semimetals should be able to incorporate Chern insulators in the sense that a general (multi-band) structure may realise both non-trivial Chern insulating phases and semimetal phases. The corresponding bulk-boundary theory should similarly encompass both chiral edge states and Fermi arcs. As we explain later, this is important in order to properly distinguish Fermi arcs topologically. Our proposal is to use (co)homology groups associated to the \emph{pair} $(T,W)$. If the Weyl surface $W$ is empty, we recover the topological insulator classification groups. General techniques from algebraic topology make these groups computable, elucidate the bulk-boundary correspondence, and reveal new rich structures in the topological semimetal theory.

The band crossings at the Fermi level occur on a submanifold $W\subset T$ and comprises several connected components $W=\coprod_{i}W_i$. The $W_i$ are not assumed to have codimension 3, so as to include the case of Weyl nodal lines in 3D. On the complement $T\setminus W$, the Bloch Hamiltonians are gapped, and there is no difficulty in defining the valence bands. Therefore, the valence bands over $T\setminus W$ can be characterised by topological invariants in $H^*(T\setminus W)$; here ${}^*$ is a natural number indicating the degree of the cohomology group, and we have dropped the $\ZZ$ in the notation since \emph{integral} cohomology is meant throughout this Letter.

The group $H^*(T\setminus W)$ contains, in a mathematically precise sense, \emph{both} the usual invariants for ordinary topological insulators and those for Weyl semimetals. Let us first describe the result for the basic situation where $T=\TT^3$ and $W$ comprises two points, i.e.\ $W_i=\{w_i\}, i=1,2$. We surround each $w_i$ with a small sphere $S_{W_i}$ and write $S_W=S_{W_1}\coprod S_{W_2}$. Then we may compute that
\begin{equation}
 H^2(\TT^3\setminus W)\cong H^2(\TT^3)\oplus H_{\rm Weyl}^2(W).\label{basicchargeinvariant}
\end{equation}
The first factor is $H^2(\TT^3)\cong\ZZ^3$, generated by Chern insulators for which the non-triviality is supported on (three independent choices of) 2D subtori in $\TT^3$. The second factor $H_{\rm Weyl}^2(W)$ is the $\ZZ$ subgroup of
\begin{equation}
H^2(S_W)=H^2(S_{W_1})\oplus H^2(S_{W_2})=\ZZ^2 \nonumber
\end{equation}
for which the chiralities of $W_i$ (which live in $H^2(S_{W_i})$) add up to zero. For $l$ Weyl points, $H_{\rm Weyl}^2(W)\equiv H_{\rm Weyl}^2(\coprod_{i=1}^l W_i)$ is the $\ZZ^{l-1}$ subgroup of $H^2(S_W)=\bigoplus_{i=1}^l H^2(S_{W_i})=\ZZ^l$ for which the chiralities sum to zero. This ``chirality cancellation'' condition resembles the Nielsen--Ninomiya theorem in lattice gauge theory which has a topological explanation \cite{NN, Witten}. We identify $H_{\rm Weyl}^2(S_W)$ as the ``Weyl charge group'' specifying the charges of the Weyl points of a semimetal. As we will explain, semimetal invariants actually live in the augmented group $H^2(\TT^3\setminus W)$, and there are inequivalent band structures with the same Weyl charges.\\

\noindent{\bf The general Weyl charge group.}
It is instructive to derive \eqref{basicchargeinvariant} in a more general context. So let $T=\TT^d$ (more generally, a connected compact $d$-manifold), and $W=\coprod_{i=1}^l W_i$ be a Weyl submanifold of dimension $m<d-1$. Cohomology groups obey a \emph{Mayer--Vietoris} (MV) principle, which is a form of locality: the cohomology of a space can be built up from the cohomology of its subspaces, with the precise details encoded in a long exact MV sequence of cohomology groups. A convenient choice is $T=(T\setminus W)\cup (D_W)$, where $D_W$ comprises disjoint open sets $D_{W_i}$ surrounding each $W_i$ (``disc bundles''). A canonical construction is to take a \emph{tubular neighbourhood} of each $W_i$ in $T$ \cite{BottTu}. Note that the boundary of $D_W$ is a \emph{sphere bundle} $S_W$ over $W$, i.e.\ a $(d-m-1)$-sphere over each point $w\in W$, and that the intersection $(T\setminus W)\cap D_W$ can be retracted onto $S_W$. Depending on how $W$ sits inside $T$, the bundle $S_W$ can be fairly complicated.

The MV sequence is an exact sequence linking the cohomology groups of the union ($T$), the subspaces ($T\setminus W$ and $D_W$), and the intersection (topologically equivalent to $S_W$). We are particularly interested in the portion which reads:
\begin{equation}
\cdots H^{d-2}(S_W)\rightarrow H^{d-1}(T)\xrightarrow{\alpha} H^{d-1}(T\setminus W) 
\xrightarrow{\beta} H^{d-1}(S_W)\xrightarrow{\Sigma} H^d(T)\cdots\;\;\;\;\;\label{MVsequence}
\end{equation}


Usually, $H^{d-1}(D_W)$ is present in the MV sequence together with $H^{d-1}(T\setminus W)$, but $D_W$ can be retracted onto $W$, so $H^{d-1}(D_W)=H^{d-1}(W)=0$ for dimensional reasons. The maps $\alpha,\beta$ in \eqref{MVsequence} are simply given by restriction (of representative differential forms). Each $S_{W_i}$ in $S_W=\coprod_i S_{W_i}$ is a connected $d-1$ dimensional manifold, so $H^{d-1}(S_{W_i})\cong \ZZ$ is a topological ``charge group'' for each $W_i$, generalising the monopole charge. Writing $l$ for the number of connected components in $W$, the map 
$$\Sigma:H^{d-1}(S_W)=\ZZ^l \rightarrow \ZZ=H^d(T)$$ 
is the MV ``connecting homomorphism'', which can be shown to be the sum of the charges in $H^{d-1}(S_W)=\bigoplus_i H^{d-1}(S_{W_i})$ \cite{BottTu}.

Exactness of \eqref{MVsequence} means that the image of one arrow is equal to the kernel of the subsequent arrow. As long as $W$ contains at least two components ($l\geq 2$), the kernel $H^{d-1}_{\rm Weyl}(W)$ of $\Sigma$, i.e.\ the ``zero-net-charge'' subgroup of $H^{d-1}(S_W)$, is nonzero; by exactness, $\beta$ maps exactly onto this subgroup, which we call the ``Weyl charge group''. We see, for example, that it is not possible to have only a single Weyl component.

The restriction map $\alpha$ takes the topological insulator group $H^{d-1}(T)$ into the total group $H^{d-1}(T\setminus W)$. Since $\beta\circ\alpha$ is zero, topological insulator invariants do not contribute Weyl charge as expected. Physically, we can imagine two energetically separated bands in an insulator coming together to form a semimetal, but such crossings can be gapped out again and are not topologically protected. Conversely, a semimetal whose band crossings can be gapped out to form an insulator does not have Weyl charges protecting the crossings.

\begin{figure}[ht]
\begin{center}
\includegraphics[scale=0.35]{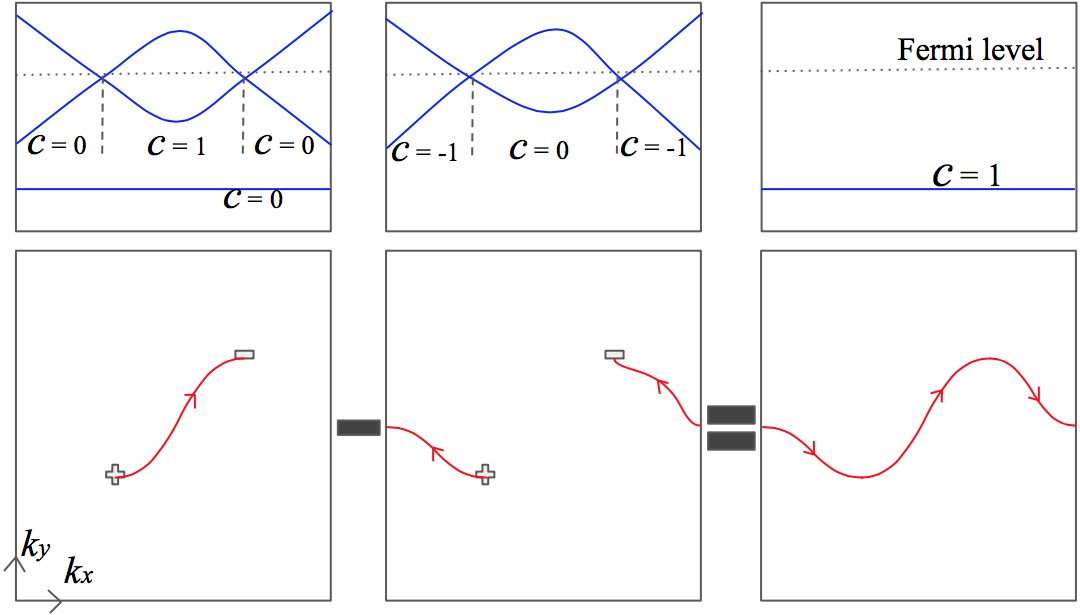}
\end{center}
\caption{\small\it The left and centre band structures (blue lines) have Weyl points with the same charges $\pm 1$. However, their bulk band structures are different as indicated by how the 2D Chern numbers $\mathcal C$ (in the $y$-$z$ direction) vary with $k_x$. Their Fermi arcs (red lines) are topologically distinct, occurring in the $k_x$ interval where the Chern numbers are non-zero. The arcs differ by a cycle of chiral edge states wrapping around the Brillouin torus in the $x$-direction. The left band structure is obtained by ``adding'' one valence band of uniform 2D Chern number 1 to the middle one.}
\label{fig:ambiguity}
\end{figure}

One might ask why it is advantageous to understand $H^{d-1}(T\setminus W)$ when the Weyl charge group $H^{d-1}_{\rm Weyl}(W)$ appears to suffice. An answer is that the argument for Fermi arc existence involves looking at (a family of) topological insulator invariants living in $d-1$ dimensional Brillouin subtori (parameterised by $k_x$ say), which involves the subgroup $H^{d-1}(T)\subset H^{d-1}(T\setminus W)$, and then invoking the (parametrised) bulk-boundary principle for topological insulators. In physical terms, it is not sufficient to simply consider the \emph{local} band topology near $W$ when analysing the bulk-boundary correspondence, i.e.\ we require \emph{global} knowledge of the semimetal band structure on $T\setminus W$. For example, we can locally deduce that a Fermi arc begins from a Weyl point with positive charge and ends at one with negative charge, but the way it wraps around the torus cycles has to be inferred from additional global information, see Fig.\ \ref{fig:ambiguity}.\\

\section{Bulk-boundary correspondence and Fermi arcs}
Some further algebraic topology machinery makes the topological nature of Fermi arcs more apparent. The bulk-boundary correspondence can be modelled by a ``topological boundary map'' \cite{PSB, MT3, MT1}, which in the simplest case can be realised by an integration map on differential forms (integrating out one direction). In the mathematics literature, such maps are referred to as ``wrong-way'' maps, Gysin maps, push-forwards, etc.. We are interested in the topological boundary map corresponding to a projection $\pi$ from $T=\TT^d$ onto a surface Brillouin zone $\wt{T}=\TT^{d-1}$. The Weyl surface projects onto $\wt{W}\subset \widetilde{T}$, and a natural home for the surface invariants is $H^{d-2}(\wt{T}\setminus\wt{W})$. We will explain how Fermi arcs naturally represent such surface invariants.

We take the topological bulk-boundary map to be the push-forward $\pi_{!}:H^{d-1}(T\setminus W)\rightarrow H^{d-2}(\wt{T}\setminus\wt{W})$, which we define via Poincar\'{e} duality (or Lefschetz/Alexander duality) \cite{Hatcher} as follows. For each $p$, there is a canonical isomorphism $H^{d-p}(T\setminus W)\cong H_p(T,W)$ where the latter is a \emph{relative homology} group, and similarly $H^{d-1-p}(\wt{T}\setminus \wt{W})\cong H_p(\wt{T},\wt{W})$. Roughly speaking, $H_p(T,W)$ is represented by $p$-chains in $T$ (maps from a $p$-simplex into $T$) whose boundaries lie in $W$. We define $\pi_{!}$ by the diagram
\begin{equation}
\label{pushforward}
\xymatrix{
H^{d-1}(T\setminus W)  \ar[d]^{\pi_{!}} \ar[rr]^{\sim\;}_{\mathrm{PD}} && H_1(T,W) \ar[d]^{\pi_*} \\
H^{d-2}(\wt{T}\setminus\wt{W})  && H_1(\wt{T},\wt{W}) \ar[ll]^{\sim}_{\mathrm{PD}} } \nonumber
\end{equation}
where $\pi_*$ is simply the composition of chains by $\pi$ (so that they map into $\wt{T}$). We remark that when $W$ is empty, we are in the situation of topological insulators, and the push-forward reduces to integration along the projection as in \cite{PSB, MT1}.
\\

\noindent{\bf Fermi arc analysis for the 3D Weyl semimetal.} We now explain how surface Fermi arcs represent elements of the group of boundary invariants $H^{d-2}(\wt{T}\setminus\wt{W})$ or, more conveniently, $H_1(\wt{T},\wt{W})$. Let us analyse $H_1(\wt{T},\wt{W})$ for the basic case of two Weyl points in $T=\TT^3$ with $\pi$ mapping $W=\{w_1,w_2\}$ to two distinct points, $\wt{W}=\wt{W}_1\coprod\wt{W}_2,\, \wt{W}_i=\{\wt{w}_i\}\equiv\{\pi(w_i)\}$, in $\wt{T}=\TT^2$. There is a homology long exact sequence whose relevant part is
\begin{equation}
  \ldots 0 \rightarrow H_1(\TT^2) \rightarrow H_1(\TT^2,\wt{W})
          \xrightarrow{\wt{\Sigma}}
H_0(\wt{W}_1\coprod\wt{W}_2) \rightarrow H_0(\TT^2). \label{homologyLES}
\end{equation}


Thus $H_1(\TT^2,\wt{W})$ contains a subgroup $H_1(\TT^2)=\ZZ^2$, generated by the two defining torus cycles. In addition, there are new non-trivial relative cycles such as paths which start at $\wt{W}_1$ and end at $\wt{W}_2$. The group $H_0(\wt{W}_1\coprod\wt{W}_2)$ is simply pairs of integers $(a_1,a_2)$ with $a_i$ assigned to $\wt{W}_i$, while $H_0(\TT^2)=\ZZ$. Since $\wt{W}_1$ and $\wt{W}_2$ are homologous in $\TT^2$ (they bound the path between them), the last arrow in $\eqref{homologyLES}$ is simply $(a_1,a_2)\mapsto a_1+a_2$. By exactness, $\wt{\Sigma}$ maps exactly onto pairs of the form $(a,-a)$. Thus, there is an extra $\ZZ$-subgroup in $H_1(\TT^2,\wt{W})$ represented by the above new relative 1-cycle (which gets mapped onto $(1,-1)$). This extra subgroup corresponds to the \emph{directed} Fermi arc from $\wt{W}_1$ to $\wt{W}_2$, and is the image of the Weyl charge subgroup $H^2_{\rm Weyl}(W)\subset H^2(\TT^3\setminus W)$ under the bulk-boundary map $\pi_!:H_1(\TT^3,W)\cong H^2(\TT^3\setminus W)\rightarrow H^1(\TT^2\setminus\wt{W})\cong H_1(\TT^2,\wt{W})$. On the other hand, the topological insulator group $H^2(\TT^3)$ is mapped under the bulk-boundary correspondence to $H_1(\TT^2)$, and therefore does not contribute to Fermi arcs.

Although there are three topologically distinct paths from $\wt{W}_1$ to $\wt{W}_2$ (two of them are shown in Fig.\ \ref{fig:ambiguity}), they differ only by some non-trivial torus cycle, i.e.\ an element in $H_1(\TT^2)$. Thus only one extra $\ZZ$ is generated by Fermi arcs, and it counts the total number of Fermi arcs. In the language of \eqref{homologyLES}, several different classes of paths in $H_1(\TT^2,\wt{W})$ are mapped onto the same element $(1,-1)$. This observation is also clear from the bulk-boundary correspondence, as illustrated in Fig.\ \ref{fig:ambiguity}. We note that the group of boundary invariants and the map $\pi_{!}$ depend sensitively on the projection direction. If $\pi$ projects onto the same point, $\wt{\Sigma}$ becomes the zero map and $H_1(\TT^2,\wt{W})\cong H_1(\TT^2)$ --- the projected Weyl points cancel and there are no Fermi arcs at all.\\

\section{Movement of Weyl points}
We can also allow the Weyl points to move around in the Brillouin zone, but in such a way that they do not come into contact throughout the motion. Formally, this means that there is a smooth ambient isotopy of the Weyl submanifold in $T$, i.e.\ there is a smooth 1-parameter family of diffeomorphisms $F_t, t\in[0,1]$ of $T$ such that the initial Weyl submanifold $W=F_0(W)=\mathrm{id}(W)$ is moved to a diffeomorphic final Weyl submanifold $W'=F_1(W)$. In particular, two Weyl points with equal and opposite charge are not allowed to coalesce to form a ``Dirac point'' --- such points are known to be gap-able in the absence of additional symmetry constraints. For the bulk-boundary correspondence, we should also consider only motions which do not line up the Weyl points along the projection $\pi$ --- this ensures that the projected Weyl points $\pi(F_t(W))$ remain separated for all $t$. It is important that the local motion of the Weyl points can be implemented globally as a smooth deformation of the global semimetal band structure. Physically, the resulting Fermi arcs should be moved around coherently with the Weyl points, and we expect that the topological invariants of the initial and final semimetals/Fermi arcs can be identified independently of the motion (isotopy-independent).

In terms of the relative homology groups represented by the initial and final Fermi arcs, there is a canonical isomorphism of the bulk semimetal invariant group $(F_1)_*:H_1(T,W)\rightarrow H_1(T,W')$ induced by the final diffeomorphism $F_1$, which is simply composition of a relative cycle in $(T,W)$ with $F_1$; similarly $(F_1)_*:H_1(\wt{T},\wt{W})\rightarrow H_1(\wt{T},\wt{W'})$ for the boundary invariant groups. Suppose the initial configuration $W$ and final configuration $F_1(W)$ of Weyl points are the same, and that $F_0=\mathrm{id}$ and $F_1$ are isotopic (via $F_t$). Then the isomorphism $(F_1)_*$ is just the identity map by the homotopy invariance of homology groups, consistent with the fact that we have simply perturbed the semimetal band structure smoothly, with the position of the Weyl points similarly perturbed.

\begin{figure}[h]
\begin{center}
\includegraphics[scale=0.29]{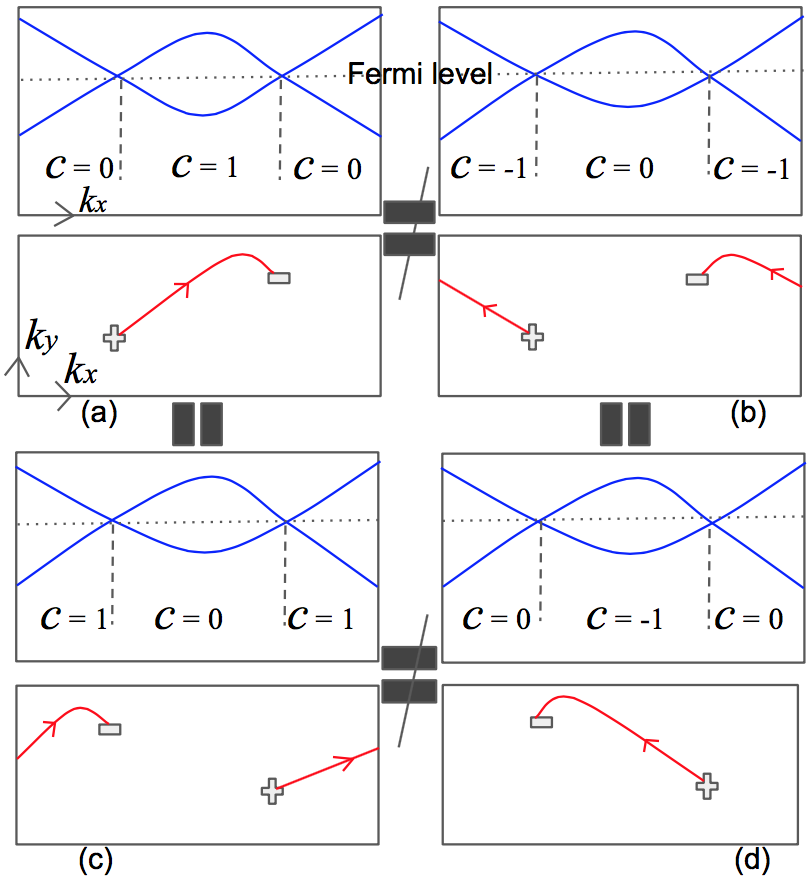}
\end{center}
\caption{\small\it Four bulk band structures (blue lines, only $k_x$ dependence shown), and their corresponding surface Fermi arcs (red lines, in $k_x$-$k_y$ coordinates) upon projecting out the $z$-direction. The local Weyl charge at a band crossing is $+$ or $-$ depending on whether the 2D Chern number $\mathcal{C}$ (in the $y$-$z$ direction) increases or decreases by 1 as $k_x$ is increased. In each of the four cases, the Fermi arcs are found in the $k_x$-region for which the bulk 2D Chern number is non-zero. The direction of the Fermi arc corresponds to the chirality (along $y$) of the edge states at each $k_x$, which is different for $\mathcal{C}=+1$ and $\mathcal{C}=-1$. Structures (a) and (c) may be identified by rotating along $k_x$ by $\pi$, and similarly for structures (b) and (d). However, (a) is inequivalent to (b), both in the bulk (different Chern numbers), and in the boundary (the directed Fermi arc cannot be moved from one to the other through globally implemented transformations of the Brillouin torus).}
\label{fig:ambiguity2}
\end{figure}

\begin{figure}[h]
\begin{center}
\includegraphics[scale=0.35]{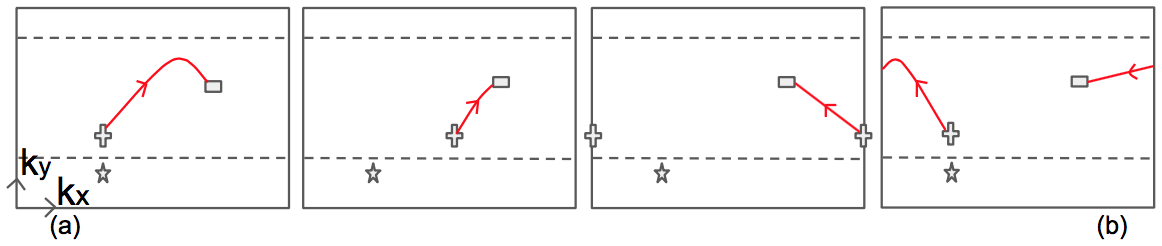}
\end{center}
\caption{\small \it The first and last diagrams are the Fermi arcs of Fig.\ \ref{fig:ambiguity2}. A star has been included in each diagram as a reference point. We can move the local patch between the dotted lines by a Dehn twist, which rotates points with lower $k_y$ by a larger angle in the $k_x$ direction. The Weyl point with charge $+1$ is moved a full round to return to its initial position, and the Fermi arc of (a) is correspondingly moved into that of (b). However, this continuous motion cannot be realised globally, i.e.\ maintaining the gluing across the lower dotted line). This can be seen from the proximity relation between the star and the Weyl point with charge $+1$, as we move through the diagrams from left to right.}
\label{fig:ambiguity3}
\end{figure}

However, if two semimetals/Fermi arcs with Weyl points $W$ are related by a diffeomorphism $F_{\rm large}$ of $T$ which is not isotopic to the identity (such as a Dehn twist), then their respective topological invariants in $H_1(T,W)$ or $H_1(\wt{T},\wt{W})$ are generally different, i.e.\ $(F_{\rm large})_*$ may not be the identity map. An illustrative example is given in Fig. \ref{fig:ambiguity2} for $T=\TT^3$ with two Weyl points. There, the bulk band structures (a) and (c) are related by moving the Weyl points (along with the Fermi arc) halfway around the $k_x$ direction. On the other hand, converting (a) to (b) requires moving the Weyl point with charge $+1$ one full round in the $k_x$ direction, while keeping the Weyl point with charge $-1$ stationary. This is essentially a Dehn twist of the Brillouin torus $T$, which is \emph{globally} obstructed from being isotopic to the identity, see Fig.\ \ref{fig:ambiguity3}. This can be seen from the fact that the bulk 2D Chern number (in the region in $k_x$ space contributing to Fermi arcs) is $+1$ in (a) but $-1$ in (b), so the bulk semimetal structures are inequivalent in a global sense. This inequivalence is also manifested in the difference in their Fermi arc topology as explained in Fig.\ {fig:ambiguity}.

The upshot is that when we wish to analyse topological stability under perturbations which move Weyl points $W\rightarrow W'$, we need to consider them in a global manner in view of the global nature of the Fermi arcs connecting the Weyl points. In analogy to the ``large gauge transformations'' in gauge theory, there exist ``large'' perturbations of semimetal band structures which move (or even fix) the Weyl points only slightly, yet result in topologically distinct phases.

One other important point is that the direction of the Fermi arc connecting one (projected) Weyl point to another reflects the chirality of the edge modes forming the arc --- this chirality is inherited from the sign of the bulk 2D Chern number. Therefore the Fermi arcs should be understood as \emph{oriented} paths, and the two Fermi arcs in Fig.\ \ref{fig:ambiguity2} (a) and (d) are topologically inequivalent both mathematically and physically. \\

The observation that diffeomorphisms which are not isotopic to the identity map can change topological invariants can be formalised mathematically as follows. The diffeomorphism group of a manifold acts by automorphisms on homotopy or (co)homology groups of the manifold, with isotopic maps having the same action. In the case of the $3$-torus, the group of connected components of (oriented) diffeomorphisms (the \emph{mapping class group}) is ${\rm SL}(3,\ZZ)$, which acts on the fundamental group $\ZZ^3\cong\pi_1(\TT^3)\cong H_1(\TT^3)$ in the natural way. For diffeomorphisms which fix the Weyl points $W$, there is similarly an action on $H_1(\TT^3,W)\cong\ZZ^4$. In the example illustrated in Fig.\ \ref{fig:ambiguity3}, we considered a Dehn twist of $\TT^3$ fixing the two Weyl points $W$, which is not isotopic to the identity on $\TT^3$ through diffeomorphisms (fixing $W$ or otherwise). This is also a Dehn twist on the projected torus $\TT^2$, which has an action on the boundary invariants $H_1(\TT^2,\wt{W})$. Consequently, the global topological invariants of two semimetals/Fermi arcs, as elements of $H^2(\TT^3\setminus W)$ or $H_1(\TT^2,\wt{W}$)), need not be the same even if their local Weyl points charges are identical.

\section*{Outlook} Similar analyses can be carried out in higher dimensions, with a ``Fermi arc group'' $\ZZ^{\wt{l}-1}$ appearing if $\wt{W}$ has $\wt{l}$-components. This group is generated by paths from $\wt{W}_i$ to $\wt{W}_{i+1}$. The ends of a Fermi arc are allowed to move around in the Weyl components, which corresponds to the possibility of continuously modifying some boundary conditions without changing the topological class of the Fermi arc. We can also analyse lower degree invariants $H^{d-p}(T\setminus W),\,p>1$, in analogy to ``weak'' topological insulator invariants, although this can be fairly hard if $W$ is complicated. Generalisation of the results in this paper have been achieved by the authors recently \cite{MT4}, where gerbes in 4D and quaternionic line bundles in 5D appear in the analysis. In the same paper, we go beyond Dirac-type Hamiltonians and introduce new classes of semimetals whose local charges are subtle Atiyah-Dupont-Thomas invariants globally constrained by the Kervaire semicharacteristic, leading to the prediction of torsion Fermi arcs.
A MV principle also holds in $K$-theory \cite{Karoubi}, which is more computable, generalises more easily to the noncommutative setting \cite{PSB, Thiang}, and has the further advantage that it directly classifies valence vector bundles up to stable isomorphism.\\


\noindent{\it Acknowledgements}. This work was supported by the Australian Research Council via ARC Discovery Project grants DP150100008 and DP130103924.


\end{document}